\documentclass[conference,a4paper]{IEEEtran}
\usepackage{cite}
\usepackage{amsmath,amssymb,amsfonts}
\usepackage{algorithmic}
\usepackage{graphicx}
\usepackage{textcomp}
\usepackage{xcolor}
\usepackage{booktabs}
\usepackage{xurl}
\usepackage[hidelinks]{hyperref}

\begin{document}

\title{Federated Lightweight Transformer for Network-Layer Cyberattack Detection in O-RAN\\
}
\author{
\IEEEauthorblockN{1\textsuperscript{st} Hong Nhung Nguyen}
\IEEEauthorblockA{
\textit{Department of Electronic Engineering} \\
\textit{Hanyang University} \\
Seoul, South Korea \\
Email: nhungnguyen.uet@gmail.com
}

\and

\IEEEauthorblockN{2\textsuperscript{nd} Abdulahi Badrudeen}
\IEEEauthorblockA{
\textit{Department of Electronic Engineering} \\
\textit{Hanyang University} \\
Seoul, South Korea\\
Email: aabadrudeen@hanyang.ac.kr
}

\and

\IEEEauthorblockN{3\textsuperscript{rd} Yekaterina Kim}
\IEEEauthorblockA{
\textit{Department of Electronic Engineering} \\
\textit{Hanyang University} \\
Seoul, South Korea\\
kattykimchi@hanyang.ac.kr
}

\and

\IEEEauthorblockN{4\textsuperscript{th} Truong Thu Huong}
\IEEEauthorblockA{
\textit{School of Electrical and Electronic Engineering} \\
\textit{Hanoi University of Science and Technology} \\
Hanoi, Vietnam \\
huong.truongthu@hust.edu.vn
}

\and

\IEEEauthorblockN{5\textsuperscript{th} Sunwoo Kim}
\IEEEauthorblockA{
\textit{Department of Electronic Engineering} \\
\textit{Hanyang University} \\
Seoul, South Korea\\
remero@hanyang.ac.kr
}
}

\maketitle
\begin{abstract}
Open Radio Access Network (O-RAN) improves radio access network flexibility through disaggregation, open interfaces, and programmable RAN Intelligent Controllers (RICs). However, the same openness and distribution increase the attack surface and make centralized intrusion detection difficult to scale across heterogeneous RAN sites. This paper presents a federated lightweight Transformer framework for privacy-preserving network-layer cyberattack detection in O-RAN. Each participating site trains a compact Transformer locally on temporally ordered network-flow records, while only model parameters are shared with a federated coordinator. To model realistic site heterogeneity, the client datasets are partitioned according to a Dirichlet distribution, resulting in non-independent and identically distributed (non-IID) local data. The local detector uses feature projection, positional encoding, multi-head self-attention, feed-forward layers, and temporal pooling to classify five network-layer attack categories: ICMP, Slowloris, SYN, TCP-ACK, and UDP. Experiments on the NetsLab 5G O-RAN intrusion-detection dataset show an overall accuracy of 94.89\% and a macro-F1 score of 81.09\%. 

\end{abstract}


\section{Introduction}
\IEEEPARstart{B}{eyond-5G} and envisioned 6G systems increasingly depend on automated control, optimization, and intelligent network management \cite{s24031038}. Open Radio Access Network (O-RAN) supports this evolution by disaggregating conventional RAN functions, introducing open interfaces, and enabling programmable control through the Non-Real-Time (Non-RT) and Near-Real-Time (Near-RT) RAN Intelligent Controllers. These capabilities create an attractive platform for artificial intelligence (AI)-driven optimization and security, but they also enlarge the attack surface across distributed cells, software components, and controller interfaces.

The authors in \cite{Polese_2023} showed that programmable RAN components and open interfaces enable data-driven network control while introducing significant interoperability, scalability, and security challenges. The authors in \cite{10495907} proposed ZTRAN, a zero-trust O-RAN security architecture in which authentication, intrusion detection, and secure slicing are implemented as xApps. These studies establish the importance of intelligent O-RAN security; however, conventional detection architectures commonly depend on centralized data collection or locally isolated models. Such approaches become less appealing when traffic records are sensitive, geographically distributed, and statistically heterogeneous.

Federated learning (FL) provides a natural mechanism for collaborative learning without centralizing raw data. Nevertheless, O-RAN deployment conditions make FL particularly challenging. Cells can differ markedly in user density, traffic composition, radio conditions, and attack prevalence, resulting in non-identically distributed (IID) client datasets. Moreover, cyberattack detection \cite{HUONG2021103509} is fundamentally temporal: abnormal behavior can emerge through a sequence of packet or flow observations rather than from a single independent record. Lightweight Transformer models are therefore promising, as self-attention enables them to capture dependencies across multiple observations while maintaining compatibility with distributed training.

This work investigates a federated lightweight transformer for network-layer attack classification in O-RAN. The main contributions are summarized as follows:
\begin{itemize}

    \item A distributed O-RAN intrusion-detection framework is formulated in which participating sites retain local network data and exchange only model parameters during federated training  (Sectionn III)
    
    \item In our work, a lightweight Transformer architecture is designed to learn short-term temporal relationships from sequences of preprocessed network-flow records. A non-IID federated environment is emulated by partitioning the training sequences across five clients using a Dirichlet distribution with a concentration parameter (Section IV)
    
\end{itemize}

The remainder of this paper is organized as follows. Section II reviews related work. Section III presents the system model and data-processing pipeline and describes the proposed federated transformer and training procedure. Section IV details the experimental setup and reports results, followed by discussion in Section V. Section VI concludes the paper.

\section{Related Work}

\subsection{AI-Based Security in O-RAN}

Machine learning has increasingly been applied to address security problems affecting O-RAN components and interfaces. Hung et al. \cite{10907911} proposed an anomaly-detection mechanism targeting malicious xApps and E2-interface attacks in the Near-RT RIC. Their results demonstrate that traffic-based learning can identify abnormal interactions between E2 nodes and the controller. However, the approach is evaluated in a centralized experimental environment and does not address collaborative learning across geographically separated RAN sites.

The disaggregated nature of O-RAN makes distributed security particularly relevant. Traffic and operational data may be generated across different radio sites and administrative domains, while privacy and operational constraints can limit raw-data sharing. As a result, an effective O-RAN security mechanism should ideally support cross-site knowledge sharing without requiring centralized collection of traffic traces.

\subsection{Federated Learning for O-RAN and Intrusion Detection}

The foundational Federated Averaging (FedAvg) algorithm proposed by McMahan et al.  \cite{6} enables multiple clients to collaboratively optimize a global model by aggregating locally trained parameters. However, statistical heterogeneity can degrade convergence when local datasets are non-IID. FedProx \cite{MLSYS2020_1f5fe839} addresses this issue by adding a proximal regularization term to the local objective, thereby constraining excessive divergence from the global model.

Several studies have investigated federated intelligence within O-RAN. Zhang et al.  \cite{10001658} developed a federated deep-reinforcement-learning framework for coordinating multiple xApps in resource allocation and network slicing. Although their study demonstrates the feasibility of federated intelligence in RIC-based architectures, its focus is network optimization rather than intrusion detection. Attanayaka et al. \cite{10278993} proposed peer-to-peer federated anomaly detection for O-RAN, allowing neighboring nodes to exchange model knowledge without depending on a single aggregation server.

Alalyan et al. \cite{10818449} proposed secure distributed FL for cyberattack detection in beyond-5G O-RAN using distributed aggregation, agent selection, and transfer learning under IID and non-IID conditions. Their work represents an important step toward decentralized O-RAN protection. However, further investigation is needed into temporal representations that can exploit relationships among consecutive traffic events.

\subsection{Transformer-Based Intrusion Detection}
Transformer architectures are well suited to sequential network-security data because self-attention can capture dependencies among observations without recurrent processing. Adjewa et al. \cite{adjewa2024efficientfederatedintrusiondetection} explored an optimized BERT-based model for federated intrusion detection in 5G networks. Their results indicate that Transformer-based models can achieve strong detection performance and that quantization can reduce deployment cost. Nevertheless, that study focuses on general 5G and IoT traffic rather than the network-layer O-RAN scenario considered here.

Realistic datasets remain another limitation. Bonati et al. \cite{Bonati_2024} introduced the Open RAN Commercial Traffic Twinning Dataset, which contains more than 500 hours of cross-layer key performance measurements and protocol-stack logs generated by replaying commercial traffic patterns on an experimental O-RAN platform. The dataset is valuable for modeling realistic O-RAN operation but primarily represents benign behavior and therefore does not provide comprehensive labels for cyberattack classification.

Overall, prior studies address O-RAN anomaly detection, federated intelligence, and Transformer-based intrusion detection largely as separate problems. This work combines these directions by studying a lightweight Transformer under non-IID federated training for multi-class network-layer cyberattack detection in an O-RAN-oriented dataset.

\section{Data Processing and System Model}
\subsection {Federated O-RAN Security Model}
Consider a set of $K$ distributed O-RAN sites participating in federated training. Each site $k$ possesses a private local dataset $D_k$ containing network-layer observations. The sites train a common intrusion-detection model locally, while a federated coordinator distributes the global parameters and aggregates local model updates. Raw network records are not transmitted to the coordinator.

At the communication round $t$, the coordinator distributes the global parameter vector $w^t$ to participating clients. Each client performs local optimization using $D_k$ and returns the updated parameter vector $w_k^{t+1}$. The coordinator then constructs the next global model from the client updates.

\subsection{Data Preparation}

The experiments use the public Network Dataset from the NetsLab 5G O-RAN intrusion-detection dataset, supported by the Network Software and Security Labs (NetsLab) \cite{Data_netlab_2025}. Before federated training, raw network-layer records are transformed into temporal sequences suitable for Transformer learning.

Numerical and categorical variables are processed separately. Numerical attributes are converted to numeric values, and missing or invalid entries are replaced using medians calculated from the training data only. Low-cardinality categorical variables, such as protocol or connection-state attributes, are encoded numerically. High-cardinality identifiers and raw textual attributes are removed because they may introduce excessive dimensionality or encourage memorization of sample-specific identifiers. The retained features are concatenated into a single feature vector and standardized using a training-fitted standard scaler. The same transformation is subsequently applied to the test data.

The dataset was divided into training and testing subsets using an 80/20 ratio.

\subsection{Proposed Federated Lightweight Transformer}

\begin{figure}[!t]
    \centering
    \IfFileExists{Fig1_FTTORS.png}{%
        \includegraphics[width=\columnwidth]{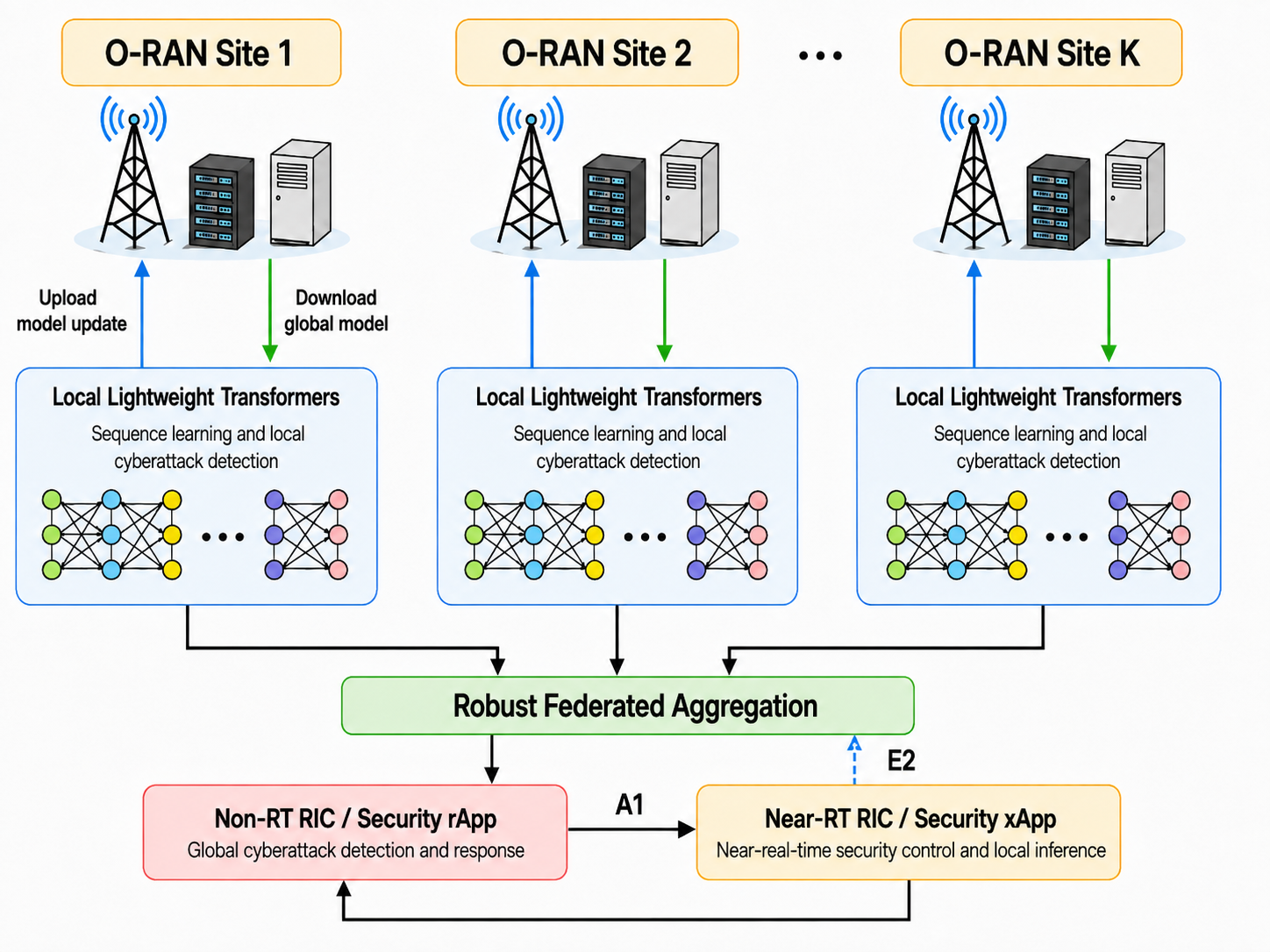}%
    }{%
        \fbox{\parbox[c][1.35in][c]{0.92\columnwidth}{\centering Placeholder for Fig1\_FTTORS.png}}%
    }
    \caption{Overview of the federated lightweight Transformer architecture for distributed O-RAN network-layer cyberattack detection.}
    \label{fig:framework}
\end{figure}

Fig.~\ref{fig:framework} illustrates the proposed framework. Each O-RAN site maintains its local traffic sequences and trains an identical lightweight transformer. At the end of a local training phase, only model parameters are transmitted to the federated coordinator. The aggregated global model is then redistributed for the next communication round. 

\subsection{Local Transformer Architecture}

Each client receives an input sequence
\begin{equation}
X \in \mathbb{R}^{L \times d},
\end{equation}
where $L=8$ the sequence length $d$ is the number of preprocessed input features. The input is projected into a 64-dimensional latent representation:
\begin{equation}
H_0 = XW_e+b_e,
\end{equation}
where $W_e$ and $b_e$ denote the learnable embedding parameters.

Because self-attention itself is permutation invariant, sinusoidal positional encoding is added to represent the order of observations within each sequence. The resulting embeddings are processed by two Transformer encoder layers. Each layer contains multi-head self-attention and a position-wise feed-forward network. Four attention heads are used.

For queries $Q$, keys $K$, and values $V$, scaled dot-product attention is defined as
\begin{equation}
\mathrm{Attention}(Q,K,V)
=\mathrm{softmax}\left(\frac{QK^{T}}{\sqrt{d_k}}\right)V,
\end{equation}
where $d_k$ is the key dimension. Multi-head attention allows the model to learn complementary temporal relationships among consecutive traffic observations.

The feed-forward subnetwork uses an intermediate dimension of 128, and dropout with probability 0.1 is applied for regularization. After the final encoder layer, the temporal representations are aggregated using mean pooling:
\begin{equation}
h=\frac{1}{L}\sum_{t=1}^{L} H_t.
\end{equation}
The pooled representation is normalized and passed to a fully connected classifier. For $C$ attack categories, the output probability vector is
\begin{equation}
\hat{y}=\mathrm{softmax}(W_ch+b_c),
\end{equation}
where $W_c$ and $b_c$ are classification parameters. The reported experiment considers five classes: ICMP, Slowloris, SYN, TCP-ACK, and UDP.

\subsection{Federated Training}
At round $t$, the client $k$ receives the global model $w^t$ and performs local training according to
\begin{equation}
w_k^{t+1}=\mathrm{LocalTrain}(w^t,D_k).
\end{equation}
local optimization using AdamW \cite{adamw} with learning rate $10^{-3}$. Each client performs two local epochs before returning its updated parameters.

The federated coordinator applies weighted FedAvg \cite{FedAver}:
\begin{equation}
w^{t+1}=\sum_{k=1}^{K}\frac{n_k}{\sum_{j=1}^{K}n_j}w_k^{t+1},
\end{equation}
where $n_k$ denotes the number of training sequences at client $k$. Five clients participate in ten communication rounds in the current configuration.

\section{Experimental Results}

\begin{table}[!t]
\caption{Experimental configuration}
\label{tab:config}
\centering
\begin{tabular}{lc}
\toprule
\textbf{Parameter} & \textbf{Value} \\
\midrule

Number of clients & $5$ \\
Dirichlet $\alpha$ & $0.5$ \\
Embedding dimension & $64$ \\
Transformer encoder layers & $2$ \\
Attention heads & $4$ \\
Feed-forward dimension & $128$ \\
Dropout & $0.1$ \\
Optimizer & AdamW \\
Learning rate & $10^{-3}$ \\
Local epochs per round & $2$ \\
Aggregation & Weighted FedAvg \\
\bottomrule
\end{tabular}
\end{table}

The final global model is evaluated on the independent test set using overall accuracy, macro-precision, macro-recall, macro-F1 score, class-wise precision, recall, F1-score, and the confusion matrix. Macro-averaged metrics are emphasized because the five attack classes are not equally represented.


The resulting overall accuracy is $94.89$ \%
The model achieved approximately 87.25\% macro-precision, 80.12\% macro-recall, and 81.09\% macro-F1. The difference between overall accuracy and macro-F1 indicates substantial variation in performance across classes.

\begin{table}[!t]
\caption{Class-Wise Detection Performance}
\label{tab:class_results}
\centering
\begin{tabular}{lccc}
\toprule
\textbf{Class} & \textbf{Precision (\%)} & \textbf{Recall (\%)} & \textbf{F1 (\%)} \\
\midrule
ICMP & $86.96$ & $32.79$ & $47.62$ \\
Slowloris & $65.94$ & $77.84$ & $71.40$ \\
SYN & $99.71$ & $97.99$ & $98.84$ \\
TCP-ACK & $98.24$ & $98.04$ & $98.14$ \\
UDP & $85.39$ & $93.95$ & $89.47$ \\
\midrule
Macro average & $87.25$ & $80.12$ & $81.09$ \\
\bottomrule
\end{tabular}
\end{table}


The strongest performance is obtained for SYN and TCP-ACK. For SYN, 6,105 of 6,230 test samples are correctly classified, corresponding to approximately $97.99$\% recall. SYN precision is approximately $99.71$\%, yielding an F1-score of approximately $98.84$\%.

Similarly, 5,408 of 5,516 TCP-ACK samples are correctly classified. This corresponds to approximately $98.04$\% recall and $98.24$\% precision, resulting in an F1-score of approximately $98.14$\%. These results indicate that the temporal network-flow characteristics of SYN and TCP-ACK are highly distinguishable under the learned Transformer representation.

For UDP, 2,051 of 2,183 samples are correctly classified, yielding approximately $93.95$\% recall. The precision reaches approximately $85.39$\%, with an F1-score of approximately $89.47$\%. The most frequent UDP error is confusion with Slowloris, with 85 UDP samples assigned to that class.

Slowloris shows moderate performance. Among 704 test samples, 548 are correctly classified, corresponding to approximately 77.84\% recall. However, 116 Slowloris samples are classified as UDP and $35$ as TCP-ACK. Precision is approximately $65.94$\%, producing an F1-score of approximately $71.40$\%.

ICMP is the most difficult class. Only $120$ a few $366$ samples are correctly identified, resulting in approximately 32.79\% recall. Precision remains comparatively high at approximately 86.96\%, but 126 ICMP samples are classified as UDP and 105 as Slowloris. The resulting ICMP F1-score is approximately $47.62$\%.

\section{Discussion}

The experimental results provide evidence that a compact Transformer can be trained collaboratively across non-IID federated clients while retaining strong overall attack-classification performance. In particular, SYN and TCP-ACK achieve F1-scores above 98\%, suggesting that these classes exhibit temporal network-flow characteristics that are consistently captured despite heterogeneous client distributions.
The results experience also demonstrates why overall accuracy should not be used as the sole evaluation measure. Although the global model achieves 94.89\% accuracy, the macro-F1 score is 81.09\%. Because the F1-score  weights each class equally, weak detection of the minority ICMP class has a substantial effect even when the majority classes are classified almost perfectly.

Potential extensions include class-weighted objectives, focal loss, balanced local sampling, client-aware aggregation, and evaluation under stronger federated heterogeneity. In addition, the present experiment uses standard weighted FedAvg. Although robust aggregation is motivated by the broader O-RAN threat model, resistance to poisoned client updates or backdoor attacks is not demonstrated by the reported results and should therefore be evaluated explicitly in future work before robustness claims are made.

Another limitation is that the current experiment addresses network-layer traffic only. O-RAN security can also involve attacks against the RIC, xApps, E2 signaling, management interfaces, and cross-layer radio telemetry. Combining network-layer flows with O-RAN-specific protocol logs and key performance measurements could provide a more complete detection model.


\begin{figure}[!t]
    \centering
    \IfFileExists{output.png}{%
        \includegraphics[width=\columnwidth]{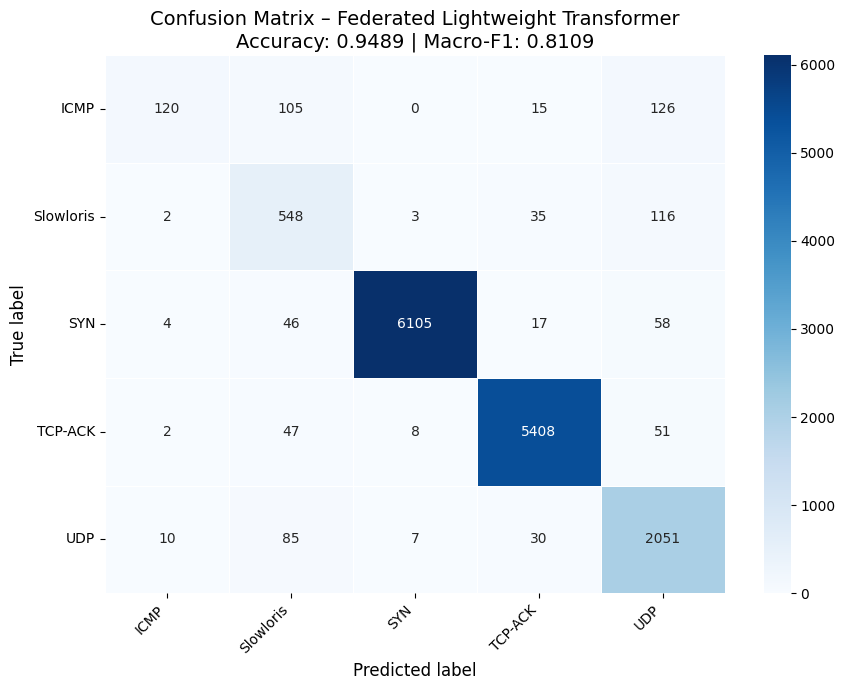}%
    }{%
        \fbox{\parbox[c][1.35in][c]{0.92\columnwidth}{\centering Placeholder for }}%
    }
    \caption{Confusion matrix of the proposed method’s output on the test dataset.}
    \label{fig:confusion}
\end{figure}

As shown in Figure \ref{fig:confusion}, the model performs well overall, especially for
Classes Slowloris, SYN, TCP-ACK, and UDP, which show high accuracy. The dominant ICMP errors involve confusion with UDP and Slowloris. 
\section{Conclusion}

This paper presented a federated lightweight transformer for distributed network-layer cyberattack detection in O-RAN. The proposed approach organizes network records into short temporal sequences and trains a compact self-attention model across five non-IID federated clients without centralizing raw traffic data. The final global model achieved 94.89\% accuracy and 81.09\% F1-score on five attack categories. SYN and TCP-ACK were detected with F1-scores above 98\%, whereas ICMP remained substantially more difficult because of low recall and confusion with UDP and Slowloris. The study shows that federated Transformer learning is a feasible approach for privacy-preserving intrusion detection across heterogeneous RAN sites. In future work, we will investigate class-balancing strategies, stronger non-IID conditions, robust aggregation under poisoned or compromised clients, communication efficiency, inference latency, and O-RAN-specific multi-layer telemetry.

\section*{Acknowledgment}
This work was supported by Institute of information \& communications Technology Planning \& Evaluation(IITP) grant funded by the Korea government(MSIT) (No. RS-2024-00397216, Development of the Upper-mid Band Extreme massive MIMO (E-MIMO))

\bibliographystyle{IEEEtran}
\bibliography{references}

\end{document}